\documentclass{article}
\usepackage{epsfig}
\usepackage{amssymb}
\usepackage{amsmath}
\usepackage{amsfonts}
\newcommand \be{\begin{eqnarray}}
\newcommand \ee{\end{eqnarray}}
\begin{document}
%\draft
%\LARGE
%\Large
%\preprint{HEP/123-qed}
%\twocolumn[\hsize\textwidth\columnwidth\hsize
%           \csname @twocolumnfalse\endcsname
%\title{Low Density Neutron Matter; Large Scattering Length}
%\title{Spin 1/2 Fermions in the Unitary Limit}
\begin{center}
%{\bf Spin 1/2 Fermions in the Unitary Limit.III}\\
{\bf Boson-Faddeev in the Unitary Limit and Efimov States}\\
\bigskip
\bigskip
H. S. K\"ohler \footnote{e-mail: kohlers@u.arizona.edu} \\
{\em Physics Department, University of Arizona, Tucson, Arizona
85721,USA}\\
\end{center}
\date{\today}
%\maketitle
%\doublespacing

\begin{abstract}
A numerical study of the Faddeev equation for bosons is made with
two-body interactions at or close to the Unitary limit. Separable
interactions are obtained from phase-shifts defined by scattering length
and effective range. In EFT-language this would correspond to NLO. 
Both ground and Efimov state energies are calculated.
For effective ranges $r_0 > 0$ and rank-1 potentials 
the total energy $E_T$ is found to converge with 
momentum cut-off $\Lambda$ for $\Lambda > \sim 10/r_0$ .
In the Unitary limit ($1/a=r_0= 0$) the energy does however diverge. 
It is shown (analytically) that in this case $E_T=E_u\Lambda^2$.
Calculations give $E_u=-0.108$ for the ground state and $E_u=-1.\times10^{-4}$ 
for the single Efimov state found.
The cut-off divergence is remedied  by modifying the  off-shell t-matrix
by replacing the rank-1 by a rank-2 phase-shift equivalent 
potential.  This is somewhat similar to the counterterm method 
suggested by Bedaque et al. 
This investigation is exploratory and does not refer to any specific
physical system.

\end{abstract}

\section{Introduction}
Systems involving particles with 2-body interactions at or close to the
Unitary limit have become of specific interest in the physics community during the
last several years for reasons that have repeatedly been pointed out in
numerous publications related to both atomic and sub-atomic problems. 
The Unitary limit is here defined as being that for which the scattering length
$a$ and effective range $r_0$ are infinite and zero respectively. In that case the
only scale left in a system of fermions of 'infinite' extension would be 
the Fermi-momentum. The total energy would then have to be proportional to
the Fermi-energy as that is the only scale left in this problem. 
Experimental results point to $E_T \sim 0.44 E_F$ with
$E_F$ being the kinetic energy of the 'unperturbed' kinetic energy of the
system. Theoretical determination of the this universal constant is a
many-body problem. It  should however only involve some simple constants
and might
\it a\' priori \rm seem straightforward. 
It has however been found to be a theoretical challenge. Some calculations
of the author using Brueckner methods show for example a very strong
dependence on the effective range $r_0$. There is also in any theoretial
calculation a necessary cut-off $\Lambda$ in momentum-space that renders
the 2-body interaction a function of this $\Lambda$.  This does introduce
another length parameter into the theory. In the previous work of the
author, and shown below, the rank-1  separable interaction is in the Unitary 
limit singular at
the momentum $\Lambda$. This fact introduces another problem; the
applicability of a many-body theory in this limit. The applicability of the
Bruckner method was for example questioned after the realisation of the
very large correlations and consequently, "wound-integral" in this case.
The "model" space represented by a zero-temperature Fermi distribution is
no longer adequate. A Green's function method including spectral broadening
would be more appropriate. 

As opposed to the 'infinite' system the three-body system is exactly solvable
by the Faddeev method \cite{fad60}.
It therefore presents a more interesting project for a numerical 
study with interactions at
and near the Unitary limit. Of specific interest here is also 
the phenomena first brought to the attention by
Vitali Efimov. \cite{eff70,ama72}  He found the surprising fact that 
 bosons interacting with a resonance in the
2-body state (i.e.  $1/a\sim 0$) would result in a strongly bound 
three-body system \it and \rm with a spectrum  of loosely bound excited states.

The inverse scattering method as applied here uses two-body on-shell data 
(scattering length and effective range) as input. Although on-shell properties of
the t-matrix are fitted exactly, many-body calculations involve also
off-shell t-matrix elements. These are not derivable from experimental
two-body data. 
\footnote{This presents a problem, of course  not restricted to the use of  
the inverse scattering method.}So even if a rank-1
potential is sufficient to fit the on-shell data as is often the case 
it leaves the off-shell
undetermined. The extension to a rank-2 will allow for a phase-shift
equivalent interaction with different off-shell t-matrix elements. This
provides a practical tool for exploring off-shell or equivalently,
three-body effects. This method will be used in some cases below.
It was used by the author in earlier work. (see e.g. \cite{hsk10})
The on-shell data relate to the asymptotic form of the two-body scattering 
wave-function.
These have to be preserved when increasing the rank which implies that  the
interaction should be modified at short range in coordinate space as shown
below.

In the present report we focus only on the ground and Efimov state
energies, as well as on the dependence on
scattering lengths and effective ranges and on the questions of convergence
as a function of cut-off in momentum-space.

In  Section 2 is found a presentation of the necessary tools which are the
Faddeev equation and the inverse scattering method.
Section 3 show results of numerical calculations in 3 subsections wth 8 Figures.
Section 4 is a summary and some discussion of the results.

\section{Formalism}
The Faddeev three-boson equation for a spin-independent rank-1 separable 
attractive potential 
 $V(k,k')=- v(k)v(k')$ is given by \begin{equation} \chi(q)=\frac{2}{{\cal I}(E_T-\frac{3}{4}q^2)}\int_{0}^{\Lambda} \frac{v(|{\bf k}+\frac{1}{2}{\bf q}|)v(|{\bf q}+\frac{1}{2}{\bf k}|)}{q^2+{\bf q}\cdot {\bf k}+k^2-E_T}\chi(q)d{\bf k}
\label{faddeev}
\end{equation}
with 
\begin{equation}
{\cal I}(s)=1+\frac{1}{2\pi^2}\int_{0}^{\Lambda}
v^{2}(k)(s-k^2)^{-1} k^{2}dk
\label{DG}
\end{equation}

The extension to  the rank-2 potentials  used below in Section 3.2 is
straight-forward. \cite{fud68,sta69}

The separable interactions are calculated from phase-shifts by an inverse
scattering method that dates back at least some 40 years. \cite{tab69}
Some recent applications by the author can be found  in refs where details can be
found.\cite{kwo95,hsk07,hsk04,hskm07,hsk09,hsk10} where some details 
of the method are also shown. 
The input are phase-shifts which in general can be either experimental 
or otherwise defined.
For the purpose of this investigation they will be defined by a scattering
length $a$ and and an effective range $r_0$.
A rank-1 separable potential is then sufficient to
reproduce the input phase-shifts exactly.  
(If the phase-shift were to 
changes sign such as in the nuclear $^1S_0$ case a rank two potential
is necessary.  \cite{kwo95}). 
%To predict the properties of a few or
%many-body system one in general also needs a three-body (and perhaps many-body)
%force. It can be absorbed in the off-shell t-matrix that can be adjusted by
%inreasing the rank in order to also fit the three-body data.
As mentioned above a higher rank may be required in order to accomodate
three-body data.
The present work is not specific to any particular system other
than that the 2-body interaction is close to the Unitary limit for which
universality would apply. It is shown however that an
off-shell correction (three-body force) via a rank-2 potential can be used to
prevent the ultraviolet divergence and collapse of the three-boson system
in that limit.

In the case of a rank-1 attractive potential one has

\begin{equation}
V(k,k')=-v(k)v(k')
\label{V}
\end{equation}
Inverse scattering then yields
(e.g. ref \cite{kwo95,tab69})

\begin{equation}
v^{2}(k)= \frac{(4\pi)^{2}}{k}sin \delta (k)|D(k^{2})|
\label{v2}
\end{equation}
where
\begin{equation}
D(k^{2})=\frac{k^2+E_B}{k^2}exp\left[\frac{2}{\pi}{\cal P}\int_{0}^{\Lambda}
\frac{k'\delta(k')}{k^{2}-k'^{2}}dk' \right]
\label{D}
\end{equation}
where ${\cal P}$ denotes the principal value  and
$\delta(k)$ are the phaseshifts. $\Lambda$ provides a cut-off in
momentum-space and the interaction is fully defined by the phase-shifts,the
two-body binding energy $E_B$ and
by $\Lambda$.  The effect of the  cut-off will be exploited below.
$E_B$ is calculated from
\begin{equation}
\sqrt(E_B)=\frac{1}{a}+\frac{1}{2}r_0E_B
\label{E_B}
\end{equation}
It is  set to zero for unbound
states. 

For the rank-2 potentials that are used 
in Sect. 3.2 the method of Chadan and Fiedeldey was
applied.\cite{cha58,fie69} (see also ref. \cite{kwo95}).
In this a set of initial phaseshifts is assumed to be
provided. An arbitrary
interaction $V_1$, is then assumed, defined either by another set of phases or
explicitly. A second potential, $V_2$ can then be constructed so that the
rank-2 potential given by $V_1$ and $V_2$ reproduce the initial phases. 
If the initial phases are, as in our case below,  all of the same sign,
they can be reproduced by a rank-1 potential from eq. (\ref{v2}). 
The extension to a rank-2
potential, allows for an arbitrary change of off-shell
behaviour or in other words of the three-body term while preserving the fit
to the initial two-body phase-shifts.

With $\delta(k)=\frac{\pi}{2}$, the unitary limit,  these phases can be
reproduced by a rank one-potential and one finds (e.g.\cite{hsk10})

\begin{equation}
v_u^{2}(k)= -\lambda \frac{(4\pi)^{2}}{(\Lambda^{2}-k^{2})^{\frac{1}{2}}}
\label{vpi2}
\end{equation}
where the factor $\lambda$ ($=1$ in the unitary limit) is introduced 
for later presentation  of results where the
three-boson binding is calculated  as a function of this strength factor.
Note that  $v_u^2(k)$ $\rightarrow -\infty$ for $k\rightarrow \Lambda$.

Note also that for $\Lambda \gg k$ one finds
\begin{equation}
v_u^{2}(k)\rightarrow -\lambda \frac{(4\pi)^{2}}{\Lambda}
\label{v2ka}
\end{equation}
In this limit, \it but only in this limit \rm ,
the unitary interaction will then be a
$\delta$-function in coordinate space with the strength  inversely
proportional to the cut-off. But eq. (\ref{vpi2}) shows that a finite
$\Lambda$ results in an abrupt increase in strength and a singularity 
at $k=\Lambda$ to
preserve the condition $\delta=\frac{\pi}{2}$ for all $k\leq \Lambda$.

With the interaction (\ref{vpi2}) and with the momenta
in units of the cut-off $\Lambda$ ($k_{\Lambda}=k/\Lambda$ and 
$q_{\Lambda}=q/\Lambda$) the Faddeev equation is:
\begin{equation}
\chi(q)=\frac{2\lambda}{{\cal I}(E_{u}(\lambda)-\frac{3}{4}
q_{\Lambda}^2)}\int_{0}^{1} 
\frac{v(|{\bf k_{\Lambda}}+\frac{1}{2}{\bf
q_{\Lambda}}|)v(|{\bf q_{\Lambda}}+\frac{1}{2}{\bf k_{\Lambda}}|)}
{q_{\Lambda}^2+{\bf q_{\Lambda}}\cdot {\bf
k_{\Lambda}}+k_{\Lambda}^2-E_{u}(\lambda)}\chi(q_{\Lambda})d{\bf k_{\Lambda}}
\label{unifadd}
\end{equation}
where $E_{u}(\lambda)=E_T/{\Lambda}^2$.

The function ${\cal I}(s)$ is after a change of variables  $[k_{\Lambda}=
sin(\theta)]$ given by
\begin{equation}
{\cal I}(s)=1+\frac{2\lambda}{\pi}{\cal P}\int_{0}^{\frac{\pi}{2}}
sin^2(\theta)(s-sin^2(\theta))^{-1} d\theta
\label{UDG}
\end{equation}

Note that any interaction for which $v^2(k)=F(k)/\Lambda$ 
would result in a similar universal equation with $E_T\propto \Lambda^2$.

It is also important to note  that with $s=k_{\Lambda}^2$ and  $\lambda=1$
one finds $${\cal I}(s)=0$$ for ${\bf ALL}$  $0<s<1$. This leads to a Reactance matrix
element $$<k|{\cal R}|k>=\frac{1}{k}tan \delta(k)\rightarrow \infty$$ i.e.
$\delta(k)=\frac{\pi}{2}$ for $0<k<\Lambda$ which is the condition for a
Unitary interaction with a cut-off $\Lambda$ in momentum-space. 

The associated resonance is of course also the origin og the Efimov-states in the
three-body system.
The number of such states were predicted to be\cite{eff70} 
\begin{equation}
N=(1/\pi)ln(|a|/r_0)
\label{N}
\end{equation}

The problem with the renormalisation of the three-boson system (i.e. the
$\Lambda^2$ divergence) was
addressed by Bedaque, Hammer and van Kolck.\cite{bed99} They resolve it by
introducing a three-body counter-term. Here this is done by extending the
rank-1 potential to a rank-2 by the method described above. The 'arbitrary'
interaction $V_1$ is defined by repulsive phases $\delta{k}=-r_c*k$, $k$ being the
relative momentum and $r_c$ a constant determined below. This simulates a
short-ranged repulsion which effectively removes the ultra-violet divergence
experienced with the rank-1 potential. The off-shell t-matrix elements in
the Faddeev equation are affected by this modification of the interaction
with results shown below.

\section{Numerical Results}
The energy of the three-boson system was calculated by solving
the Faddeev equation numerically either by iteration or by
matrix diagonalisation in the conventional
fashion. The separable interaction was obtained by the inverse scattering
method referred to above for a range of scattering lengths and effective
ranges including the Unitary limit,
In the results presented below the energies are
in units of $\hbar ^2/m$ and the lengths in units of $fm$, but are in
general universal.

\subsection{Ground States with rank-1 potential}
\begin{figure}
\centerline{
\psfig{figure=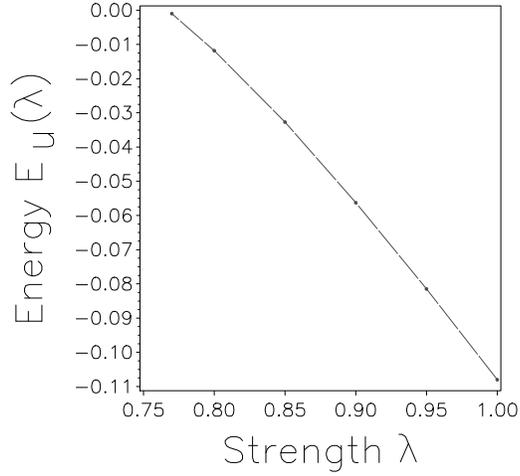,width=7cm,angle=0}
}
\vspace{.0in}
\caption{The energy $E_u$, defined in the text,  
as a function of the strength $\lambda$
of the Unitary interaction  (\ref{vpi2}). The energy of the three-boson system
would be $E_T=-E_u(\lambda)\times \Lambda^2$.  The calculated values are 
indicated by points and connected by lines for clarity.
}
\label{eff6}
\end{figure}

It was verified numerically that the 
quantity $E_u(\lambda)$, defined above, is independent of the cut-off $\Lambda$.
The dependence of $\lambda$ for the ground state is shown in Fig. \ref{eff6}
with
$E_u(\lambda=1)=-0.108$ (i.e. the Unitary limit) while $E_u=0$ for
$\lambda \sim 0.77$ with a nearly linear dependence on $\lambda$.

\begin{figure}
\centerline{
\psfig{figure=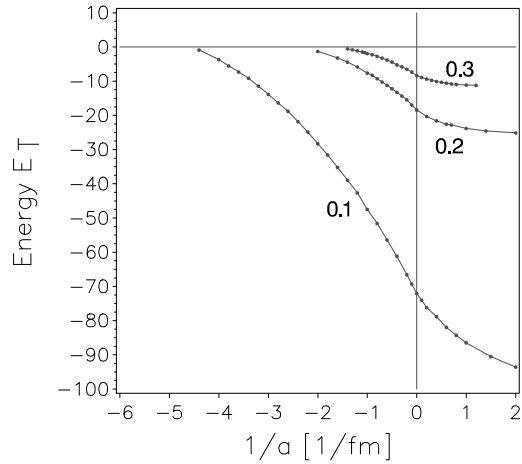,width=7cm,angle=0}
}
\vspace{.0in}
\caption{The energy of the three-boson system is shown as a function of the
scattering length and three effective ranges that are the  parameters defining
the 2-body interaction as
described in the text. The calculated values are indicated by points and
connected by lines for clarity.
}
\label{eff}
\end{figure}

Fig. \ref{eff} shows the energy as a function of $1/a$ for three chosen values of
effective ranges $r_0$. One notices a very strong dependence on the effective
range. 

\begin{figure}
\centerline{
\psfig{figure=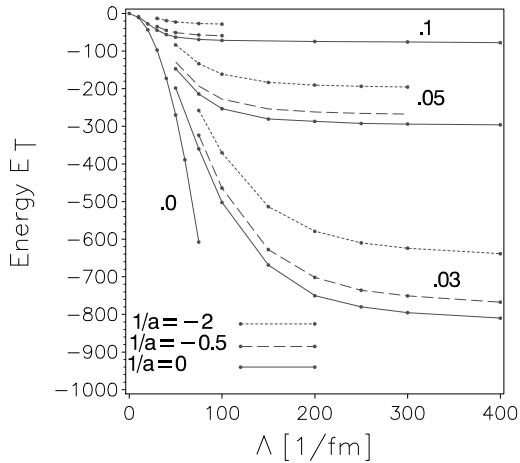,width=7cm,angle=0}
}
\vspace{.0in}
\caption{ The  energy $E_T$ of the three-boson system is shown as a function of
momentum cut-off $\Lambda$ for three
different values of scattering lenghts and four different effective ranges
as indicated. See text for further discussion.
 The calculated values are indicated by points and connected by lines for clarity.
}
\label{eff5}
\end{figure}

Fig. \ref{eff5} shows energies $E_T$ as a function of cut-off $\Lambda$ for
$1/a=0$ and some selected values of effective ranges $r_0$. It was
shown above that for $1/a=0$ and $r_0=0$, i.e. the Unitary limit ,
$E_T =0.108\Lambda^2$. This is shown numerically by the curve labelled ".0".
The figure shows that the convergence with $\Lambda$ improves 
as the effective-range is increased. The
value of $\Lambda=\Lambda_c$ at which the asymptotic value is reached
scales roughly as $10/r_0$.

\begin{figure}
\centerline{
\psfig{figure=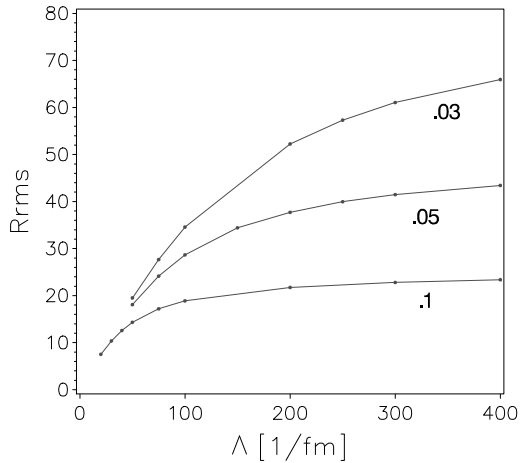,width=7cm,angle=0}
}
\vspace{.0in}
\caption{The 'size' of the three-boson system in momentum-space is shown as a
function of cut-off $\Lambda$ for the same effective ranges as in Fig.
\ref{eff5}. Only the results for $1/a=0$ are shown. Note the similar
$\Lambda$-dependence as is exhibited in Fig. \ref{eff5}.
 The calculated values are indicated by points and 
 connected by lines for clarity.
}
\label{eff7}
\end{figure}

The size of the three-boson system in momentum-space  scales roughly the same,
consistent with that the size in coordinate space would be $\approx r_0$.
Fig. \ref{eff7} shows the rms radius $R_{rms}$ of the $\xi(q)$ function. It follows

closely the $\Lambda$-dependence of  $E_T$ shown in Fig. \ref{eff5}.
\footnote{One may however note a slower approach to the asymptotic value at large
$\Lambda$. This is a general characteristic of any energy vs size display.}
One concludes that the size of the system (in momentum-space) determines
the minimum range $\Lambda_c$ of momenta that the interaction has to span. 
This is analogous to the situation  found in nuclear matter  Brueckner 
calculations
where the minimum cut-off is found to be $2k_f$ i.e. twice the
fermi-momemtum. \cite{hsk04}
In the present work on the three-boson system we also find,
quite naturally, that $R_{rms}$ is inversely proportinal to $r_0$. This is
a difference from the Brueckner calculations where the effect of
correlations on the momentum distribution is ignored and approximated by
the non-interacting fermi-distribution. This is an approximation related to the
quasi-particle approximation which is implicit in the Brueckner method.
The Green's function approach goes beyond this approximation including the
finite width of the spectral-functions and the momentum-distribution is
then wider.

\subsection{Rank-2 potential in Unitary Limit}
\begin{figure}
\centerline{
\psfig{figure=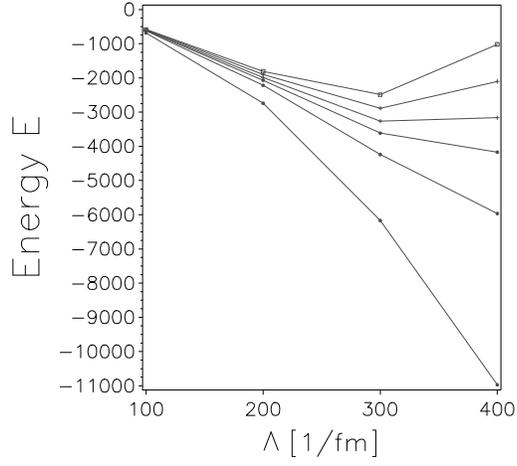,width=7cm,angle=0}
}
\vspace{.0in}
\caption{The three-boson energy as a fuction of $\Lambda$ for different
values of $r_c$. From bottom and up:
$r_c=0,.001,.0012,.0013,.0014.,0015$ [$fm$].
The calculated values are indicated by dots, connected by straight lines.
}
\label{eff22}
\end{figure}

The renormalisation of the non-relativstic three-body problem with short
ranged forces was addressed by Bedaque et al\cite{bed99}. As shown above,
the three-boson system with the two-body unitary interaction (\ref{vpi2})
collapses as $\Lambda^2$. Bedaque et al suggests to introduce a three-body
counterterm for the renormalisation. One may alternatively choose to
introduce a similar counterterm by changing the off-shell two-body t-matrix.
As alreday announced in Sect 2 this is done by replacing the rank-1 by a
rank-2 interaction with
$V_1$ chosen so as to modify the short-ranged, ultraviolet, part of the
interaction in order to prevent the collapse. With  $\delta{k}=-r_c*k$ the
related potential $V_1$ is obtained by inverse scattering. Fig. \ref{eff22}
shows the three-boson energy as a function of cut-off $\Lambda$ for three
different values of $r_c$. The lowest curve is for $r_c=0$ i.e. the rank-1
potential and shows the $\Lambda^2$ divergence. The upper five curves are
for increasing values of $r_c$ as shown in the figure caption. A drastic
change is seen with $r_c=0.12$ showing  convergence for large $\Lambda$.  

\subsection{Efimov States}
\begin{figure}
\centerline{
\psfig{figure=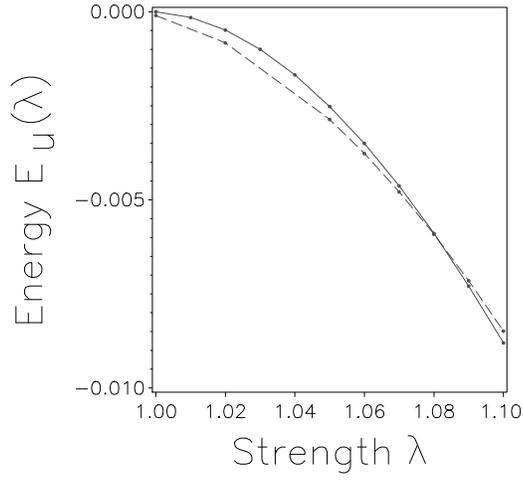,width=7cm,angle=0}
}
\vspace{.0in}
\caption{The broken line shows the Efimov state as a function of the strength
$\lambda$ of the unitary interaction (\ref{vpi2}). The energy $E_u$ is given
in units of $\Lambda^2$ as in eq. (\ref{unifadd}). The full line shows the
two-body bound state energy. 
%The broken line that lies below this full line
%represents Efimov states. Any states above the full line represent two
%bound bosons and one free. 
}
\label{eff81}
\end{figure}

There are many publications related to the low-bound excited states found
as solutions of the Faddeev equations,  states first discoverd
by  and named after Vitali Efimov\cite{eff70}
Numerous calculations were done at and close to the unitary limit.
Never was found more than one state that could be identified as an 'Efimov'
state although eq. (\ref{N}) suggests several states should be found with
$1/a=r_0 \sim 0$. The search for these states  was in general very elusive and 
in particular very sensitive to
very small changes in the low momenta part of the interactions. 
The broken curve in Fig. \ref{eff81}  shows the excited state energy
$E_u(\lambda)$ obtained from solving eq. (\ref{unifadd}). The solid line
represents thw two-body bound state energy. The sole Efimov state is seen
to coincide with the two-body at $\lambda=1.08$. The Faddeev equation also
yields numerous three-boson energies located above the two-body curve. 
These represent dimers, two bound bosons, and a free boson. The ground state
energies  are some fifty times deeper.

\begin{figure}
\centerline{
\psfig{figure=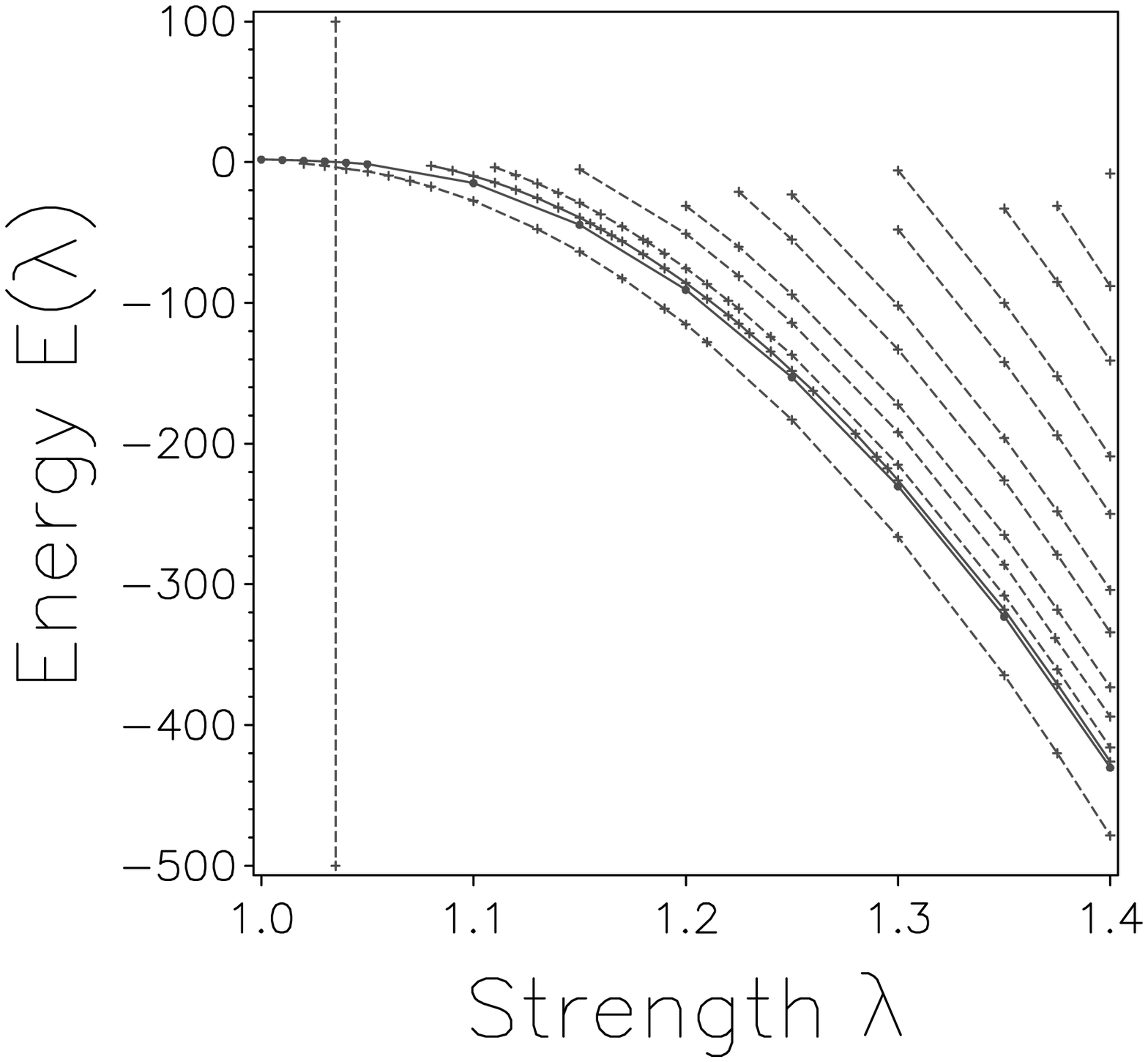,width=7cm,angle=0}
}
\vspace{.0in}
\caption{ Energies as a function of the strength $\lambda$.
The solid line shows the two-boson bound
state energies.
The sole line below this is the Efimov state. The numerous
lines above are dimer+1 states and would not be classified as Efimov
states.
}
\label{eff11}
\end{figure}

Another example is shown in Fig. \ref{eff11}. The separable interaction is
here defined by  a rank-1 potential with scattering length $a=-2$ 
and an effective range
$r_0=.03$. Solutions of the Faddeev equation are shown as a function of the
strength multiplier $\lambda$. The solid line shows the two-boson bound
state energies.(Not clearly shown is that they are unbound for
$\lambda<1.04).$ 
The sole line below this is the Efimov state. The numerous
lines above are dimer+1 states. 

\begin{figure}
\centerline{
\psfig{figure=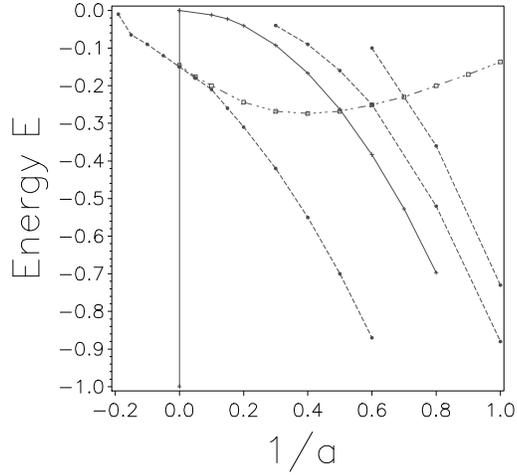,width=7cm,angle=0}
}
\vspace{.0in}
\caption{The energy as a function of $1/a$ ($a$ is scattering length) with
$r_0=0.1$. The solid line shows the dimer bound state energy. The broken
lines below this line show  Efimov state energies, while those above are
dimer+1 energies. The cut-off parameter is here $\Lambda=75$. (Cf Fig.
\ref{eff5}). See text for further details.
int
}
\label{eff20}
\end{figure}

Yet another example of the numerous calculations that were made 
is shown in Fig. \ref{eff20}.
The effective range is here $r_0=0.1$ and the scattering length is allowed
to vary over the range indicated. The dots connected with broken lines
indicate solutions of Faddeev equations with the binding energy $E_B$ from
eq. (\ref{E_B}). According to other works these should with incresing $1/a$
approach the bound dimer line. This is not the case here. This may be a
characteristic of the separable interaction. Humberston et al \cite{hum68}
show a comparison of separable (non-local) and local (Yukawa and
exponential) interactions used in three-boson calculations. The energy
increases much faster with the strength of the interaction for the
separable than it does for the local interaction. 
In order to investigate the effect of the binding,  another set
of calculations were made shown by the broken line connecting squares. Here
$E_B$ is set to zero in eq. (\ref{D}). It does have a $1/a$ dependence similar
to what was expected from the literature. 
Only one Efimov state is found here with $r_0=0.1$. From eq. (\ref{N}) 2 states
could be expected only for $1/a<.025$.
It should be mentioned that expression (\ref{N}) was tested by Huber \cite{hub85}.
Our result may be associated with the rank-1 potential.

\section{Summary and Comments}
Fig. \ref{eff} shows that the energy of the bound system of three-bosons  depend
strongly on the range $r_0$ of the two-body interaction. This is consistent with
earlier results for the infinite fermion-system.\cite{hsk07,hsk08}
Fig. \ref{eff5} shows that as a function of the momentum cut-off 
$\Lambda$, the energies converge toward 
asymptotic values $E_T$ that, as in Fig.\ref{eff}, are
functions of the range $.03<r_0<.1$ but largely independent of the scattering
length for $|a| \gg r_0$. Convergence was reached at 
$\Lambda_c \sim 10/r_0 \sim 5\times R_{rms}$.  The size of the trimer at
equilibrium (saturation), the inverse of $R_{rms}$ scales as $r_0$ 
and the range $\Lambda$ has to scale with $R_{rms}$ consistent with our
results. 

For comparison the curve labelled by ".0" in Fig. \ref{eff5} shows the 
energy vs $\Lambda$ in the unitary limit. As shown above (eq.
(\ref{unifadd})) this limit has to be treated as a special case 
giving the analytic result  $E_T=E_u(\lambda) \Lambda ^2$. 
The numerical result shown by Fig. \ref{eff6} is $E_u(\lambda=1)=-0.108$ 
while $E_u(\lambda)=0$ for 
$\lambda \sim 0.77$.

The quadratic divergence and simaltaneous collapse of the boson trimer in 
the Unitary limit was dampend by a renormalisation procedure consisting 
of a counter-term represented by a rank-2 potential with results shown in
Fig. \ref{eff22}. 

Efimov states were identified although not quite as expected which may be a
consquence of the specific interactions.


\begin{thebibliography}{10}
\bibitem{fad60} L.D. Faddeev, Zh. Eksperim. Teor. Fiz. {\bf 39} (1960)
                1459;
                Sov. Phys. JETP(transl.) {\bf 12} (1961) 1014.
\bibitem{eff70}  V. Efimov,
                 Phys. Lett. {\bf B33}  (1970) 903.
\bibitem{ama72}  H.D. Amado and J.V. Noble,
                 Phys. \ Rev. D {\bf 5} (1972) 1992.
\bibitem{fud68} M.G. Fuda,
		 Nucl.Phys. {\bf A116} (1968) 83.
\bibitem{sta69} R.W. Stagat, Nucl.Phys. {\bf A125} (1969) 654.
\bibitem{tab69}  Frank Tabakin,
                Phys. Rev. {\bf 177} (1969) 1443.
\bibitem{kwo95}  N.H. Kwong and H.S. K\"ohler, 
                 Phys. \ Rev. C {\bf 55} (1997) 1650.
\bibitem{hsk07}  H.S. K\"ohler ,
                 nucl-th/0705.0944.
\bibitem{hsk04}  H.S. K\"ohler ,
                nucl-th/0511030.
\bibitem{hsk09}  H.S. K\"ohler ,
                 nucl-th/0907.1539.
\bibitem{hsk10}  H.S. K\"ohler ,
                 nucl-th/1008.3884.
\bibitem{cha58}  K. Chadan, Nuovo Cimento {\bf 10} (1958) 892; Nuovo
                 Cimento A {\bf 47} (1967) 510.
\bibitem{fie69}  H. Fiedeldey, Nucl.Phys. A {\bf 135} (1969) 353.
\bibitem{hskm07} H.S. K\"ohler and S.A. Moszkowki,
                 nucl-th/0703093.
\bibitem{hsk08}  H.S. K\"ohler ,
                 nucl-th/0801.1123.
\bibitem{bed99}  P.F. Bedaque, H.-W. Hammer,U. van Kolck,
                 Nucl. Phys. {\bf A646} (1999) 444.
\bibitem{hum68}  J.W. Humberston, R.L. Hall and T.A.Osborn
                 Phys. Lett. {\bf 27B}  (1968) 195.
\bibitem{hub85}  Stephen Huber,
                 Phys. \ Rev. a {\bf 31} (1985) 3981.

\end{thebibliography}
\end{document}